\documentclass[journal]{IEEEtran}

\usepackage{graphicx}
\usepackage{cite}
\usepackage{siunitx}
\DeclareSIUnit{\arbunit}{a.u.}   % displacement unit; reported in arbitrary units
\usepackage{microtype}
\usepackage{amsmath}
\usepackage{amssymb}

\usepackage{amsthm}
\usepackage{booktabs}
\usepackage{algorithm}
\usepackage{algpseudocode}
\usepackage{enumitem}
\usepackage{tikz}
\usepackage{tikzscale}
\usetikzlibrary{shapes,arrows,calc,fit}
\usepackage{pgfplots}
\usepackage{eso-pic}
\pgfplotsset{compat=newest}

\definecolor{matblue1}{rgb}{0,0.4470,0.7410}
\definecolor{matred1}{rgb}{0.85,0.325,0.098}
\definecolor{matyel1}{rgb}{0.9290, 0.6940, 0.1250}
\definecolor{matpur1}{rgb}{0.4940, 0.1840, 0.5560}
\definecolor{matgre1}{rgb}{0.4660, 0.6740, 0.1880}

\definecolor{mplblue}{HTML}{1F77B4}
\definecolor{mplorange}{HTML}{FF7F0E}
\definecolor{mplgreen}{HTML}{2CA02C}
\definecolor{mplred}{HTML}{D62728}
\definecolor{mplpurple}{HTML}{9467BD}
\definecolor{mplbrown}{HTML}{8C564B}
\definecolor{mplpink}{HTML}{E377C2}
\definecolor{mplgray}{HTML}{7F7F7F}
\definecolor{mplolive}{HTML}{BCBD22}
\definecolor{mplcyan}{HTML}{17BECF}

\newtheorem{remark}{Remark}

\newcommand{\legsymwidth}{5mm}   % length of the drawn line segment
\newcommand{\legsymlw}{1.5pt}    % line width of the segment

\NewDocumentCommand{\legsym}{ m O{solid} O{} }{%
  \tikz[baseline=-0.6ex]{%
    \legsymDrawLine{#1}{#2}%
    \legsymDrawMark{#1}{#3}%
  }%
}
\newcommand{\legsymDrawLine}[2]{%
  \def\legsymTmp{#2}\def\legsymNone{none}%
  \ifx\legsymTmp\legsymNone\else
    \draw[#1,#2,line width=\legsymlw] (0,0) -- (\legsymwidth,0);%
  \fi
}
\newcommand{\legsymDrawMark}[2]{%
  \def\legsymTmp{#2}\def\legsymEmpty{}%
  \ifx\legsymTmp\legsymEmpty\else
    \coordinate (lsm) at ($(0,0)!0.5!(\legsymwidth,0)$);%
    \def\legsymT{circle}\ifx\legsymTmp\legsymT   \filldraw[#1] (lsm) circle[radius=2.1pt];\fi
    \def\legsymT{ocircle}\ifx\legsymTmp\legsymT  \draw[#1,fill=white,line width=0.9pt] (lsm) circle[radius=2pt];\fi
    \def\legsymT{square}\ifx\legsymTmp\legsymT   \filldraw[#1] ($(lsm)+(-1.9pt,-1.9pt)$) rectangle ($(lsm)+(1.9pt,1.9pt)$);\fi
    \def\legsymT{osquare}\ifx\legsymTmp\legsymT  \draw[#1,fill=white,line width=0.9pt] ($(lsm)+(-1.9pt,-1.9pt)$) rectangle ($(lsm)+(1.9pt,1.9pt)$);\fi
    \def\legsymT{diamond}\ifx\legsymTmp\legsymT  \filldraw[#1] ($(lsm)+(0,2.5pt)$) -- ($(lsm)+(2.5pt,0)$) -- ($(lsm)+(0,-2.5pt)$) -- ($(lsm)+(-2.5pt,0)$) -- cycle;\fi
    \def\legsymT{odiamond}\ifx\legsymTmp\legsymT \draw[#1,fill=white,line width=0.9pt] ($(lsm)+(0,2.5pt)$) -- ($(lsm)+(2.5pt,0)$) -- ($(lsm)+(0,-2.5pt)$) -- ($(lsm)+(-2.5pt,0)$) -- cycle;\fi
    \def\legsymT{triangle}\ifx\legsymTmp\legsymT \filldraw[#1] ($(lsm)+(0,2.4pt)$) -- ($(lsm)+(-2.2pt,-1.7pt)$) -- ($(lsm)+(2.2pt,-1.7pt)$) -- cycle;\fi
    \def\legsymT{otriangle}\ifx\legsymTmp\legsymT \draw[#1,fill=white,line width=0.9pt] ($(lsm)+(0,2.4pt)$) -- ($(lsm)+(-2.2pt,-1.7pt)$) -- ($(lsm)+(2.2pt,-1.7pt)$) -- cycle;\fi
    \def\legsymT{x}\ifx\legsymTmp\legsymT        \draw[#1,line width=1.3pt] ($(lsm)+(-2pt,-2pt)$) -- ($(lsm)+(2pt,2pt)$) ($(lsm)+(-2pt,2pt)$) -- ($(lsm)+(2pt,-2pt)$);\fi
    \def\legsymT{plus}\ifx\legsymTmp\legsymT     \draw[#1,line width=1.3pt] ($(lsm)+(-2.3pt,0)$) -- ($(lsm)+(2.3pt,0)$) ($(lsm)+(0,-2.3pt)$) -- ($(lsm)+(0,2.3pt)$);\fi
  \fi
}

\AddToShipoutPictureBG*{%
    \AtPageUpperLeft{%
        \setlength\unitlength{1in}%
        \hspace*{\dimexpr0.5\paperwidth\relax}%%
        \makebox(0,-0.8)[c]{%
            \begin{tabular}{c}
                J.S. van Hulst \emph{et al.}, ``Precision Specimen Positioning in Electron Microscopy through Hysteresis Compensation,\\
                Iterative Learning, and Vision-Based Sensing.'' This work has been submitted to the IEEE for possible publication.\\
                Copyright may be transferred without notice, after which this version may no longer be accessible.
            \end{tabular}%
        }%
    }%
}%

\AddToShipoutPictureBG*{%
\AtPageUpperLeft{%
\setlength\unitlength{1in}%
\hspace*{\dimexpr0.5\paperwidth\relax}
\makebox(0,-21.3)[c]{
\footnotesize
\begin{tabular}{c c}
\copyright~The Authors
\end{tabular}}}}

\begin{document}

\title{Precision Specimen Positioning in Electron Microscopy through Hysteresis Compensation, Iterative Learning, and Vision-Based Sensing}

\author{J.S.~van~Hulst,~A.M.C.~de~Peffer,~D.~Herceg,~E.M.~Franken,~E.~Verschueren,~W.P.M.H.~Heemels,~and~D.J.~Antunes% <-this % stops a space
\thanks{This work is part of the research project entitled \emph{Learning in Motion}, a collaboration between the Eindhoven University of Technology and several industry partners. This project is co-financed by Holland High Tech, top sector High-Tech Systems and Materials, with a PPP innovation subsidy for public-private partnerships for research and development.}% <-this % stops a space
\thanks{J.S. van Hulst, A.M.C. de Peffer, D. Herceg, W.P.M.H. Heemels, and D.J. Antunes are with the Department of Mechanical Engineering, Control Systems Technology, Eindhoven University of Technology, PO Box 513, 5600MB Eindhoven, the Netherlands (e-mail: j.s.v.hulst@tue.nl).}% <-this % stops a space
\thanks{E.M. Franken and E. Verschueren are with Thermo Fisher Scientific, Eindhoven, the Netherlands.}}

\maketitle

\begin{abstract}
    Electron microscopy requires nanometer-scale specimen positioning over a long stroke. Piezo-stepper actuators are well suited for this task, but their accuracy is limited by hysteresis, mechanical misalignments, and non-collocated sensing. Prior work has addressed these limitations on simplified lab setups. However, extending to a full electron microscope stage introduces coupled nonlinear kinematics and, importantly, the absence of a dedicated point-of-interest (POI) sensor. This paper presents an integrated feedforward framework for precision positioning on such a stage inside an operational electron microscope. Per-element hysteresis compensation first linearizes the actuator response. In the absence of a dedicated POI sensor, a POI measurement is constructed from EM images through cross-correlation-based image tracking. From this measurement, we construct an encoder-based proxy for the POI position. Commutation-angle-domain iterative learning control then uses this proxy as its error signal to cancel the repeatable disturbances of stepping. Because the learned corrections are parameterized in the commutation angle, they transfer across the quasi-static range of drive frequencies. The framework reduces the POI tracking error by over 13$\times$ on the lab setup and by 7 to 12$\times$ on an operational transmission electron microscope.
\end{abstract}

\section{Introduction}\label{sec:intro}
\IEEEPARstart{E}{lectron} microscopy (EM) requires nano-scale positioning of a specimen over a long stroke to enable high-resolution imaging~\cite{Williams2009}. An illustrative example of this is electron tomography, where the stage is tilted through a series of angles to reconstruct the three-dimensional structure of the specimen. The electron beam rarely passes exactly through the axis of rotation, so without correction the specimen drifts out of the field of view as the stage rotates. The stage must reposition the specimen at each tilt angle to hold it at the point of interest, and any residual error introduces artifacts into the reconstruction. The quality of the resulting images therefore hinges on positioning accuracy.

Piezo-stepper actuators (PSAs) are well suited for this task because they combine high stiffness with a large range of motion~\cite{Li2005,Fleming2014}. Several parasitic effects, however, limit their positioning accuracy. Hysteresis in the piezoelectric material causes a nonlinear, history-dependent relationship between applied voltage and displacement. Mechanical misalignments arising from manufacturing tolerances introduce repeatable, cyclic errors during stepping. Non-collocated sensing, where encoders are located at the actuators rather than at the specimen, compounds these effects because quasi-static structural deformations cause discrepancies between the measured and true positions. Tilt-induced coupling between the axes further complicates positioning during tomographic acquisition.

Piezoelectric hysteresis has been studied extensively. Classical physics-based models such as those of Preisach, Prandtl--Ishlinskii, and Duhem provide analytical modeling frameworks~\cite{Visintin1994,Krejci1996}. These models range from rate-independent formulations, which depend only on the input history, to rate-dependent ones that also account for the velocity of the applied voltage. Recent data-driven approaches tailored to piezo-stepper stages apply hysteresis compensation~\cite{Strijbosch2023} and piezo waveform optimization~\cite{Aarnoudse2023}. Iterative Learning Control (ILC) has proven effective for canceling repeatable disturbances in precision positioning systems~\cite{Bristow2006,Ahn2007}. The combination of hysteresis compensation and ILC is particularly powerful for piezo-stepper stages, where hysteresis and misalignment effects are often entangled~\cite{VanMeer2025a}.

Recently,~\cite{VanMeer2025a} demonstrated that combining rate-dependent hysteresis compensation with commutation-angle-domain ILC can reduce the RMSD of a single piezo-stepper to \SI{8.7}{\nano\meter} at a drive frequency of \SI{2}{\hertz}. Although this represents a fifteenfold improvement compared to traditional feedforward, the study relied on a simplified 1DOF lab setup with a single actuator, trivial kinematics, and a direct capacitive sensor at the point of interest (POI), providing an idealized environment for feedforward design and validation.

Extending these results to the full EM stage presents several additional challenges. The stage has four degrees of freedom: three translational piezo-stepper axes and one rotational axis. Nonlinear kinematics relate these to the 3D POI position. No direct position sensor is available at the POI inside the microscope. Structural deformation and tilt-induced coupling further cause the encoder-based estimate to differ from the true POI position. Image-based motion estimation is well established in electron microscopy, for example for correcting beam-induced specimen motion in cryogenic EM acquisition~\cite{Zheng2017}. In particular, cross-correlation-based shift estimation can generate a POI position signal from EM images, as demonstrated in~\cite{VanHorssen2020,VanHulst2026}. In~\cite{VanHorssen2020}, this signal was further used for feedback at specimen level to counteract thermal drift. The central question addressed here is how to use such image-based observations for feedforward learning in an operational electron microscope, without a dedicated POI sensor.

This paper presents an integrated framework for precision specimen positioning in an operational electron microscope, extending the methods of~\cite{VanMeer2025a} from a simplified single-PSA lab setup with a dedicated POI sensor to a real EM stage where no such sensor is available. The core new capability is to learn the feedforward corrections against a POI ground truth reconstructed from EM images, so that the same learning procedure runs with or without a dedicated sensor. The contributions are as follows.
\begin{itemize}
    \item A commutation-angle-dependent sensor deviation is learned from the available POI measurement in each operating environment. On the lab setup, we use the capacitive reading $p_c$, and on the EM, we use the vision-based position estimate $p_s$. The deviation signal corrects encoder readings for quasi-static bending. The corrected estimate $\hat{p}$ provides a high-rate POI proxy that drives feedforward learning on both platforms.
    \item The commutation-angle-domain ILC of~\cite{VanMeer2025a} is extended from one PSA to three, with per-PSA error signals obtained from the corrected POI estimate $\hat{p}$ through the inverse kinematics. This transformation is identified from data and is non-trivial on the EM, where the measurement observes only the in-plane POI motion. We also reduce the hysteresis compensation to a two-parameter affine function of the history variable $h$, a simplification of the rate-dependent radial-basis-function model of~\cite{VanMeer2025a}.
    \item All methods are validated on a full 4-degree-of-freedom lab setup and on an operational transmission electron microscope.
\end{itemize}

The remainder of the paper is organized as follows. Section~\ref{sec:system} describes the positioning stage, the coordinate systems, and the problem definition. Section~\ref{sec:estimation} introduces the two sensing modalities and the sensor deviation. Section~\ref{sec:control} presents the control design, covering hysteresis compensation and ILC. Section~\ref{sec:experiments} shows experimental results and Section~\ref{sec:conclusion} concludes the paper.

\section{System Description and Problem Definition}\label{sec:system}
The positioning stage is studied in two environments: inside the electron microscope, and on a lab setup that permits a direct measurement at the POI.

\subsection{Positioning Stage}\label{sec:setup}
\begin{figure}[t!]
    \centering
    \includegraphics[width=\linewidth]{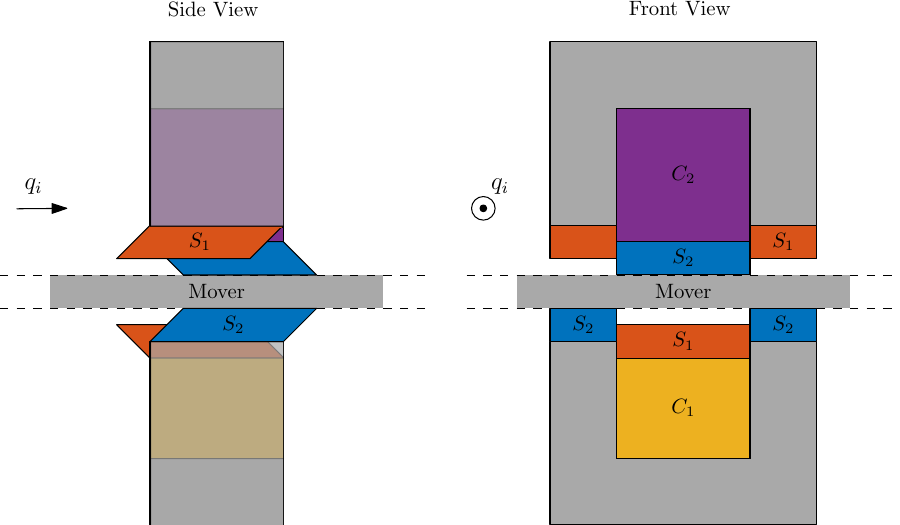}
    \caption{Schematic of a piezo-stepper actuator, following~\cite{VanMeer2025a}. The clamps ($C_1$, $C_2$) press the shear elements ($S_1$, $S_2$) onto the mover. When a shear element is in contact with the mover, it expands or contracts laterally to push or pull the mover in the $q_i$ direction.}
    \label{fig:piezo_scheme}
\end{figure}

\begin{figure}[t!]
    \centering
    \includegraphics[width=\linewidth]{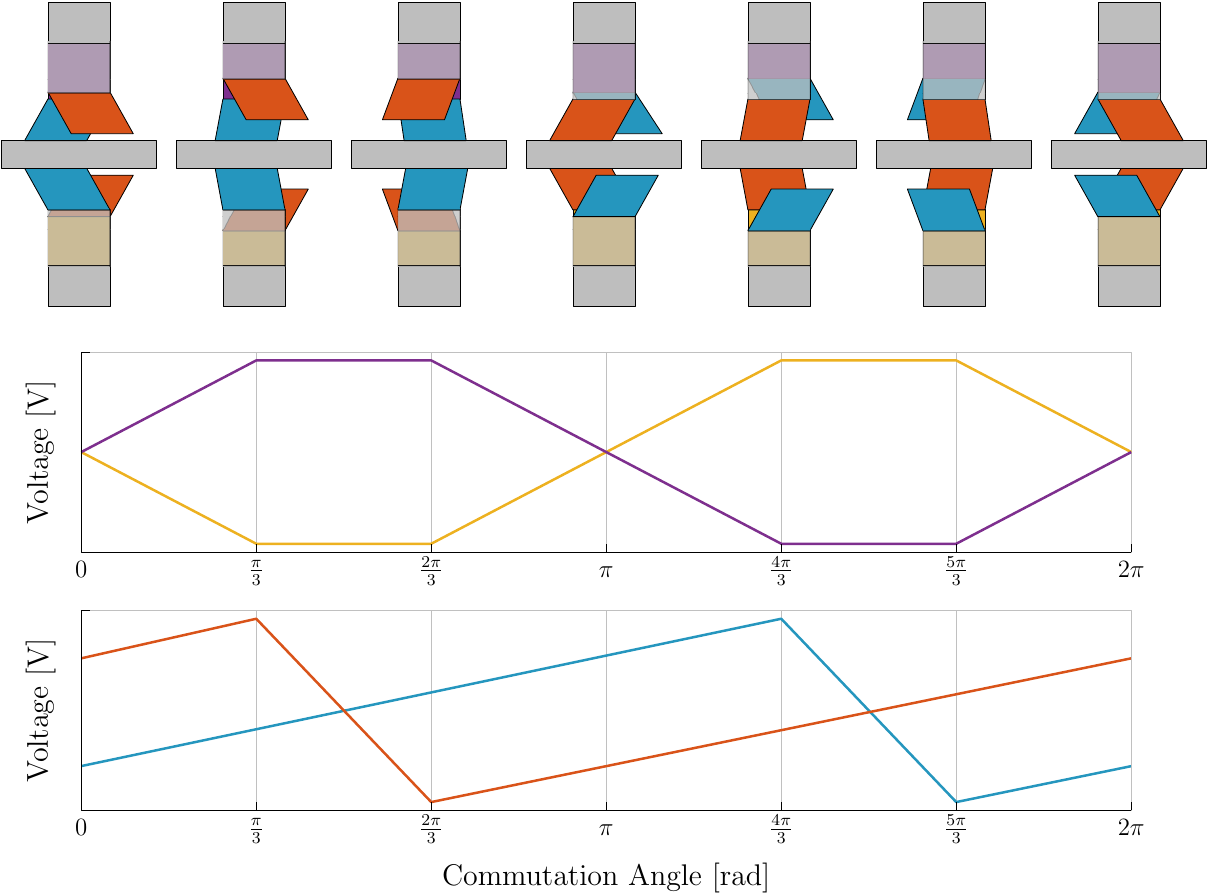}
    \caption{Voltage waveforms for piezo elements to achieve a stepping motion, following~\cite{VanMeer2025a}. The clamps (\protect\legsym{matyel1}[solid],\protect\legsym{matpur1}[solid]) press the shears (\protect\legsym{matblue1}[solid],\protect\legsym{matred1}[solid]) onto the mover one by one, and the shears drag the mover along in the lateral direction.}
    \label{fig:waveforms}
\end{figure}

The full EM positioning stage consists of three piezo-stepper actuators ($i \in \{1,2,3\}$) and one rotational actuator providing a tilt angle $\varphi \in \mathbb{R}$ that rotates the full stage. Each piezo-stepper contains two pairs of clamp and shear piezoelectric elements ($C_1, S_1, C_2, S_2$). A stepping cycle is parameterized by a commutation angle $\alpha_i \in [0, 2\pi)$: in the first half-cycle, clamp $C_2$ presses shear $S_2$ onto the mover, which expands laterally to displace the mover. When $S_2$ reaches its end of stroke, $C_1$ engages $S_1$ and $C_2$ releases, creating a continuous walking motion with a net advance of approximately \SIrange{3}{4}{\micro\meter} per full cycle. Figures~\ref{fig:piezo_scheme} and~\ref{fig:waveforms} illustrate the piezo-stepper architecture and the commutation-angle-dependent voltages applied to the elements.

\begin{figure*}[t!]
    \centering
    \includegraphics[scale=1.2]{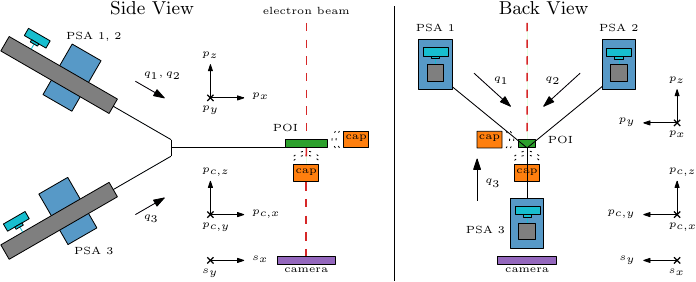}
    \caption{Schematic of the motion stage for both the lab setup and the electron microscope. The capacitive POI sensors (\protect\legsym{mplorange}[none][square]) are present only in the lab setup and measure the POI through $p_c$, while the electron beam (\protect\legsym{mplred}[dashed]) and the EM camera (\protect\legsym{mplpurple}[none][square]) are present only in the microscope and measure the POI through $p_s$. The three piezo-stepper axes $q_1$, $q_2$, $q_3$ are shown, along with the POI position $p$.}
    \label{fig:setup}
\end{figure*}

Figure~\ref{fig:setup} illustrates the positioning stage in both operating environments. For model learning and validation, a separate 4DOF lab setup with the exact same geometry is used, which allows placement of capacitive sensors at the POI. These sensors provide a direct POI measurement, denoted $p_c \in \mathbb{R}^3$. No such sensor is available during EM operation. Instead, the microscope acquires images at a rate $F_v = F_s / N_v$, where $F_s$ is the encoder sampling rate and $N_v$ is the number of frames between vision updates. The value of $N_v$ depends on the camera model and acquisition settings and is typically in the range of tens to several hundred. These images can be processed to obtain in-plane specimen shift estimates, as detailed in Section~\ref{sec:vision}. This absence of a direct POI sensor inside the microscope is one of the central challenges in the present work.

Two position variables appear throughout this paper. The encoder measurements $q = [q_1, q_2, q_3]^\top \in \mathbb{R}^3$ are linear displacements recorded at the three PSAs, which are imperfect observations of the true PSA positions subject to sensor noise. The POI coordinates $p = [p_x, p_y, p_z]^\top \in \mathbb{R}^3$ are the Cartesian position of the specimen at the point of interest. The mapping between these two quantities is discussed in Section~\ref{sec:fk}.

\subsection{Coordinate Transformations}\label{sec:fk}
A nonlinear forward kinematics model relates the encoder readings to an estimated POI position:
\begin{equation}\label{eq:fk}
    \hat{p}_0 = \Phi(q, \varphi).
\end{equation}
Here, $\Phi$ is a kinematic model, and $\varphi$ is the tilt angle. The notation $\hat{p}_0$ emphasizes that this is a model-based estimate, not the true POI position $p$. Model errors, noise, and dynamic effects cause a discrepancy between $\hat{p}_0$ and $p$. The corresponding inverse kinematics model is denoted
\begin{equation}\label{eq:ik}
    \hat{q} = \Gamma(p, \varphi),
\end{equation}
and maps a desired POI position to the piezo coordinates needed to reach it. The inverse kinematics is used later for projecting POI errors onto individual piezo coordinates (Section~\ref{sec:ilc}).

The kinematic maps $\Phi$ and $\Gamma$ capture the nominal rigid-body geometry of the stage. The true POI position deviates from this prediction for several reasons. Contact forces during commutation elastically deform the mover, producing a quasi-static encoder--POI offset that is repeatable in the commutation angle $\alpha$. Structural flexibility, differential thermal expansion, and tilt-dependent coupling cause further deviations that are not $\alpha$-periodic. The present work targets only the $\alpha$-periodic offset, which is corrected by the sensor deviation of Section~\ref{sec:enc_est}.

\subsection{Problem Definition}\label{sec:problem}
The objective is to minimize the POI tracking error $e_p(k) = r_p(k) - p(k)$, where $k$ is the discrete-time sample index, $r_p \in \mathbb{R}^3$ is the desired POI trajectory, and $p$ is the true POI position. Since $p$ is not directly measurable inside the microscope, a POI estimate $p_e$ is used in its place, whose form depends on the operating environment and is detailed in Section~\ref{sec:estimation}. Because each PSA is actuated and learned individually, the error is assessed in piezo coordinates, where the reference and the estimate are mapped through the inverse kinematics and compared per axis,
\begin{equation}\label{eq:proj}
    \varepsilon_i(k) = [\Gamma(r_p(k), \varphi)]_i - [\Gamma(p_e(k), \varphi)]_i, \quad i \in \{1,2,3\}.
\end{equation}
The kinematic map $\Phi$ is invertible on the stage workspace for each fixed $\varphi$, so $\Gamma$ is a bijection and $\varepsilon_i = 0$ for all $i$ exactly when $p_e = r_p$. Driving these per-axis errors to zero is therefore equivalent to nulling the POI error. Performance is then quantified per axis by the root-mean-square deviation
\begin{equation}\label{eq:rmsd}
    \sigma_i = \sqrt{\frac{1}{N} \sum_{k=1}^{N} \varepsilon_i(k)^2}.
\end{equation}

\section{POI Estimation}
\label{sec:estimation}
Iterative learning control requires a POI error signal at the encoder rate $F_s$. The encoders provide a signal at this rate, but contact forces during commutation elastically deform the mover and bias the reading relative to the true specimen position. Correcting this bias requires a POI ground-truth measurement against which the correction can be learned. On the lab setup this ground truth is the capacitive reading $p_c$. On the EM, where no such sensor exists, it is a vision-based estimate $p_s$ derived from the EM images. Section~\ref{sec:vision} describes how $p_s$ is obtained, Section~\ref{sec:proj_fit} how the kinematic maps are calibrated, and Section~\ref{sec:enc_est} how the learned correction turns the encoder reading into a high-rate POI proxy $\hat{p}$.

\subsection{Vision-Based Position Sensing}\label{sec:vision}
EM images provide an independent source of in-plane POI information. A cross-correlation-based tracker compares consecutive images and estimates the in-plane specimen displacement~\cite{VanHorssen2020,VanHulst2026}. The tracker runs on GPU, enabling processing at the camera frame rate~\cite{VanHulst2026}. The output is a 2D inter-frame shift
\begin{equation}\label{eq:vision}
    s(k) = [s_x(k),\, s_y(k)]^\top \in \mathbb{R}^2,
\end{equation}
where $k$ indexes camera frames and $s_x(k)$, $s_y(k)$ are noisy measurements of the in-plane specimen displacement in the $p_x$ and $p_y$ directions since the previous frame. The tracker is sampled at $F_v = F_s / N_v$, substantially slower than the encoders.

Converting the pixel-level shift to physical length units depends on the optical settings of the microscope. Assuming parallel illumination is used, the dominant parameter is the selected magnification value. A large inter-frame shift broadens the correlation peak and can produce outliers. The specimen motions considered here are slow enough that the shift remains small and the tracker produces reliable estimates.

How well $s(k)$ covers the motion of each piezo axis depends on $\varphi$, because $\Phi$ couples the piezo axes to the in-plane and out-of-plane directions through the tilt. At $\varphi = 0$, all three axes contribute to the out-of-plane direction $p_z$, so $s(k)$ captures only a fraction of each axis's motion. At $\varphi \approx \pi/6$~rad, PSA 2's axis lies entirely in the imaging plane and its full motion is observed by $s(k)$. At $\varphi \approx -\pi/6$~rad, PSA 1's axis lies entirely in the imaging plane instead. During calibration, the tilt angle can be set to maximize the coverage of the axis under calibration. During tomographic acquisition, $\varphi$ is prescribed by the imaging protocol and the coverage of each axis changes accordingly.

The specimen position relative to a reference frame $k_0$ is estimated by accumulating consecutive inter-frame shifts. Because the tracker observes only in-plane motion, the accumulation updates the two in-plane coordinates while the out-of-plane coordinate is held fixed:
\begin{equation}\label{eq:ps}
    p_s(k) = p_s(k_0) + \sum_{j=k_0+1}^{k} \begin{bmatrix} s(j) \\ 0 \end{bmatrix} \in \mathbb{R}^3.
\end{equation}
The reference $p_s(k_0)$ is initialized from the encoder-based forward-kinematics estimate, $p_s(k_0) = \Phi(q(k_0),\, \varphi)$, which anchors $p_s$ to an absolute POI position rather than to an arbitrary origin. Between camera frames, $p_s$ is held at its most recent value. The resulting estimate $p_s \in \mathbb{R}^3$ serves as the POI ground truth for learning the sensor deviation, as described in the next section.

\subsection{Kinematic Calibration}\label{sec:proj_fit}
In practice, the kinematic maps are obtained by calibration. The experiments cover a small region of the workspace, over which $\Phi$ is well approximated by its linearization at the current tilt angle. Let $\bar{p}$ denote the available POI measurement, with $\bar{p} = p_c$ on the lab setup and $\bar{p} = p_s$ on the EM. Since only the in-plane components of $p_s$ are measured~\eqref{eq:ps}, the EM fit uses those two components. To identify the linearization, a calibration move displaces all three PSAs while $\bar{p}$ and $q$ are recorded. Expressing both signals relative to the start of the move at sample $k_0$ removes the constant offset between the coordinate frames, and the linearized forward map follows from the least-squares fit
\begin{equation}\label{eq:kfit}
    \hat{K} = \arg\min_{K} \sum_{k=1}^{N_K} \big\| \bar{p}(k) - \bar{p}(k_0) - K \big(q(k) - q(k_0)\big) \big\|_2^2,
\end{equation}
where $N_K$ is the number of recorded samples. The fit yields $\hat{K} \in \mathbb{R}^{3 \times 3}$ on the lab setup and $\hat{K} \in \mathbb{R}^{2 \times 3}$ on the EM.

On the lab setup, $\hat{K}$ is invertible, and $[\Gamma(\bar{p}, \varphi)]_i$ is evaluated by applying the $i$-th row of $\hat{K}^{-1}$ to $\bar{p} - \bar{p}(k_0)$ and adding $q_i(k_0)$. On the EM, $\hat{K}$ has no inverse: two measured components cannot separate three piezo coordinates. However, in all experiments only one PSA steps at a time, and its motion appears in the measurement along the corresponding column $\kappa_i$ of $\hat{K}$. The piezo coordinate is therefore recovered by projecting onto this column, with the row vector $\kappa_i^\top / \|\kappa_i\|_2^2$ taking the place of the $i$-th row of $\hat{K}^{-1}$. This recovery is exact for motion of PSA $i$ alone, and simultaneous stepping of several PSAs would require an out-of-plane measurement (Section~\ref{sec:vision}).

\subsection{Encoder-Based POI Estimation}
\label{sec:enc_est}
\begin{figure}[t!]
    \centering
    \includegraphics[width=0.55\linewidth]{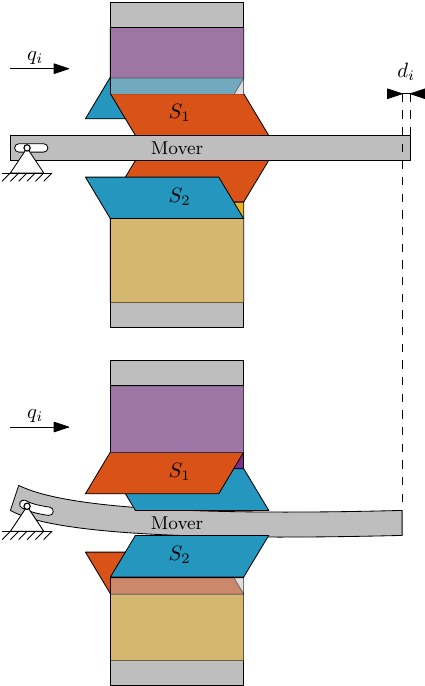}
    \caption{Illustration of the bending effect during clamp engagement. Contact forces between the clamp and shear elements cause local deformation that shortens the effective length of the mover. Because the encoder is located at the PSA, it does not capture this deformation, resulting in a commutation-angle-dependent discrepancy between the measured and true positions.}
    \label{fig:bending}
\end{figure}

\begin{figure}[t!]
    \centering
    \includegraphics[width=\linewidth]{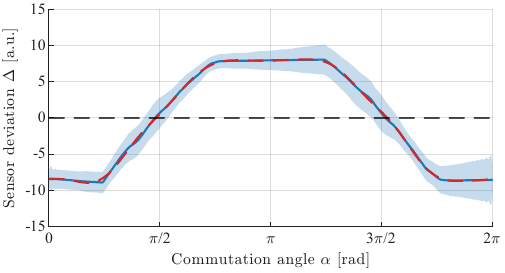}
    \caption{Mean projected encoder--POI residual $d_i$~\eqref{eq:residual} (\protect\legsym{mplblue}[solid]) and learned piecewise-linear sensor deviation $\Delta_i(\alpha_i)$~\eqref{eq:offset_param} (\protect\legsym{mplred}[dashed]) versus commutation angle $\alpha_i$ for one PSA. The shaded area denotes $\pm\sigma$ of the residual samples.}
    \label{fig:residual_alpha}
\end{figure}

Applying the forward kinematics~\eqref{eq:fk} to the encoder readings gives the baseline estimate $\hat{p}_0 = \Phi(q, \varphi)$, available at the full encoder rate $F_s$. This estimate, however, does not account for a clamping-induced bending effect that arises during commutation. When a clamp engages the mover, the resulting contact force causes a small deformation of the mover, as illustrated in Figure~\ref{fig:bending}. Because the encoder is not located at the end of the mover, it does not capture this deformation, and a discrepancy between the encoder reading and the true POI position arises. The effect is quasi-static: it depends on the current commutation angle but not on the stepping velocity, as confirmed by the data in Figure~\ref{fig:residual_alpha}.

To correct for this discrepancy, a sensor deviation is applied in piezo coordinates before computing the forward kinematics. Because the deformation depends on the commutation angle, the deviation is parameterized as a function of $\alpha$:
\begin{equation}\label{eq:phat}
    \hat{p} = \Phi\!\big(q + \Delta(\alpha),\, \varphi\big),
\end{equation}
where $\Delta(\alpha) = [\Delta_1(\alpha_1),\, \Delta_2(\alpha_2),\, \Delta_3(\alpha_3)]^\top \in \mathbb{R}^3$ collects three scalar deviation functions, one per PSA. Each $\Delta_i : [0, 2\pi) \to \mathbb{R}$ is a piecewise-linear function stored in a lookup table that maps the commutation angle of PSA $i$ to a scalar correction of its encoder reading. These scalar functions are obtained by projecting the POI estimation residual onto the individual piezo axes through the inverse kinematics $\Gamma$~\eqref{eq:ik}, as detailed below. The decomposition into independent per-PSA contributions is supported by the physical structure of the stage: the three movers are mechanically separate, so contact forces on PSA $i$ do not propagate a bending deformation to the other two. Correcting $q$ rather than $p$ also keeps the correction scalar per PSA, whereas a correction in POI space would require a three-component vector.

The sensor deviation is learned from the POI measurement $\bar{p}$. For each PSA $i$, the discrepancy between $\bar{p}$ and the encoder reading is recorded over multiple commutation cycles while PSA $i$ is stepping and the others are stationary. Although $p_s$ is sampled more slowly than the encoder, the recording spans many commutation cycles, so the samples cover the full $\alpha$ grid. Transforming the POI residual to piezo axis $i$ through the inverse kinematics $\Gamma$~\eqref{eq:ik} gives the raw residual
\begin{equation}\label{eq:residual}
    d_i(k) = [\Gamma(\bar{p}(k),\, \varphi)]_i - q_i(k).
\end{equation}
The sensor deviation is parameterized as a linear combination of basis functions,
\begin{equation}\label{eq:offset_param}
    \Delta_i(\alpha_i) = \boldsymbol{\psi}_\Delta^\top(\alpha_i)\, \boldsymbol{\theta}_{\Delta,i},
\end{equation}
where $\boldsymbol{\psi}_\Delta(\alpha_i) \in \mathbb{R}^{n_\Delta}$ collects $n_\Delta$ basis functions and $\boldsymbol{\theta}_{\Delta,i} \in \mathbb{R}^{n_\Delta}$ are the coefficients. In practice, piecewise-linear basis functions on $n_\Delta$ equidistant grid points in $[0,2\pi)$ are used for simplicity and interpretability. Stacking the basis function evaluations over all $N_\Delta$ recorded samples gives the matrix $\boldsymbol{\Psi}_\Delta = [\boldsymbol{\psi}_\Delta(\alpha_i(1)), \ldots, \boldsymbol{\psi}_\Delta(\alpha_i(N_\Delta))]^\top \in \mathbb{R}^{N_\Delta \times n_\Delta}$, and stacking the residual samples gives $\mathbf{d}_i \in \mathbb{R}^{N_\Delta}$. The coefficients are obtained by minimizing the squared residual:
\begin{equation}\label{eq:learn_o}
    \boldsymbol{\theta}_{\Delta,i} = \arg\min_{\boldsymbol{\theta} \in \mathbb{R}^{n_\Delta}} \sum_{k=1}^{N_\Delta} \big(d_i(k) - \boldsymbol{\psi}_\Delta^\top(\alpha_i(k))\, \boldsymbol{\theta}\big)^2.
\end{equation}
Because the model is linear in the parameters, the minimization reduces to a standard linear least squares problem with the closed-form solution
\begin{equation}\label{eq:lsq_solution}
    \boldsymbol{\theta}_{\Delta,i} = (\boldsymbol{\Psi}_\Delta^\top \boldsymbol{\Psi}_\Delta)^{-1} \boldsymbol{\Psi}_\Delta^\top \mathbf{d}_i.
\end{equation}
Figure~\ref{fig:residual_alpha} shows the raw residual and the learned fit for a representative PSA on the lab setup.

Including the sensor deviation reduces the residual $\bar{p} - \hat{p}$ significantly compared to the baseline $\bar{p} - \hat{p}_0$, confirming that the quasi-static bending is the dominant source of discrepancy. The remaining residual is attributed to unmodeled effects such as flexible-mode vibrations, which are not captured by a static correction.

\section{Feedforward Control Design}\label{sec:control}
\begin{figure}[t!]
    \centering
    \includegraphics[width=\linewidth]{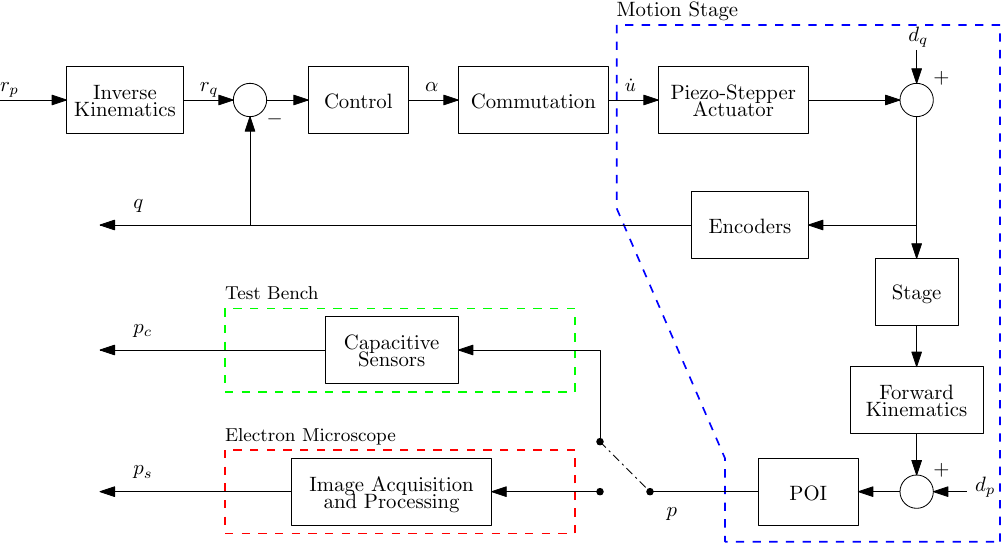}
    \caption{Block diagram showing the signal flow of the system. A POI reference $r_p$ is fed into the inverse kinematics $\Gamma$ to obtain the piezo reference $r_q$. This reference is tracked by low-level controllers. The PSAs obtain voltage signals from the commutation block, which can include hysteresis compensation and ILC-modified waveforms. The encoder signal $q$ provides a measurement of the piezo positions, while the POI position $p$ is either measured directly by the capacitive sensors $p_c$ in the lab setup, or estimated from the EM images as $p_s$ in the microscope.}
    \label{fig:block_diagram}
\end{figure}

Figure~\ref{fig:block_diagram} shows a block diagram of the control architecture. In the diagram, $d_q$ denotes disturbances at the actuator level that are reflected in the encoder reading $q$, such as misalignment and handover transients, while $d_p$ denotes disturbances between the actuator and the POI that the encoder does not capture, such as clamp-induced bending, thermal drift, and flexible-mode vibration. The control proposed in this work is entirely feedforward and acts within the commutation block, which generates the voltages applied to the piezoelectric elements. Two modifications are made there. Per-element hysteresis compensation (Section~\ref{sec:hysteresis}) inverts the voltage-to-displacement nonlinearity of each element. Commutation-angle-domain ILC (Section~\ref{sec:ilc}) then modifies the nominal commutation waveforms of Figure~\ref{fig:waveforms} to cancel the repeatable, $\alpha$-periodic disturbances of stepping. Of the disturbances above, the feedforward cancels only the $\alpha$-periodic parts, namely misalignment and handover transients through the ILC and clamp-induced bending through the sensor deviation. Both components are calibrated offline.

\subsection{Per-Piezo Hysteresis Compensation}\label{sec:hysteresis}
Hysteresis in the piezoelectric material causes a nonlinear, history-dependent relationship between applied voltage and element displacement. The waveform generator must invert this relationship to translate desired displacements into applied voltages accurately. This inversion is a prerequisite for the ILC of Section~\ref{sec:ilc}. The ILC corrects disturbances that are periodic in the commutation angle $\alpha$, but uncompensated hysteresis introduces errors that depend on the voltage history rather than on $\alpha$. These history-dependent errors are not $\alpha$-periodic and therefore corrupt the ILC error signal~\cite{VanMeer2025a}.

The compensation is applied independently to each of the three PSAs, where each PSA contains multiple piezoelectric elements. Following~\cite{VanMeer2025a}, we treat the clamp elements separately per movement direction and treat the shear elements direction-independently. Every element has the same architecture, so the model structure and identification procedure are identical for all elements. The equations below are written for a single generic element, so PSA and element indices are omitted for simplicity.

The voltage-displacement relation of an element traces a loop rather than a single curve. Such loops are captured by a range of models, from the operator-based Preisach~\cite{Visintin1994} and Prandtl--Ishlinskii~\cite{Krejci1996} models to algebraic descriptions such as the Ramberg--Osgood model~\cite{Ramberg1943}. For a piezo-stepper the loop is set by the most recent reversal, so the relevant history is the excursion
\begin{equation}\label{eq:history_t}
    h(t) = |u(t) - u(\bar{t})|
\end{equation}
of the voltage since that reversal at time $\bar{t}$. Following the memory-element model of~\cite{Strijbosch2023}, the displacement rate is the voltage rate scaled by an incremental gain that depends on this history,
\begin{equation}\label{eq:ct_hyst}
    \dot{y} = M(h)\,\dot{u},
\end{equation}
with element displacement $y$, applied voltage $u$, and incremental gain $M > 0$. This is a special case of that memory-element model, which on each branch coincides with the Ramberg--Osgood hysteresis.

The gain cannot be measured directly, as the individual element displacements are not sensed. The element current stands in for the displacement, since the displacement rate is proportional to the current, $\dot{y} = \xi\,\iota$, with $\xi$ an unknown constant~\cite{Fang2013}. Used in the incremental model~\eqref{eq:ct_hyst}, the ratio of current to voltage rate follows the gain up to $\xi$,
\begin{equation}\label{eq:proxy}
    m := \left|\frac{\iota}{\dot{u}}\right| \approx \frac{M}{\xi},
\end{equation}
with per-sample values $m(k)$. The proxy uses only the element's own voltage and current, so each element is identified on its own and no position measurement is needed.

\begin{figure*}[t!]
    \centering
    \begin{minipage}{0.32\textwidth}
        \centering
        \includegraphics[width=\textwidth]{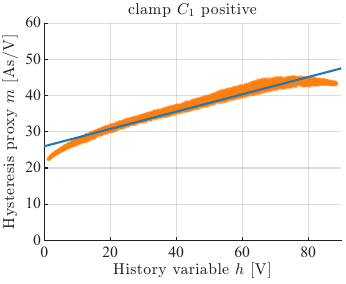}
    \end{minipage}
    \begin{minipage}{0.32\textwidth}
        \centering
        \includegraphics[width=\textwidth]{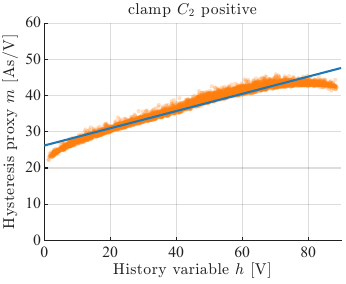}
    \end{minipage}
    \begin{minipage}{0.32\textwidth}
        \centering
        \includegraphics[width=\textwidth]{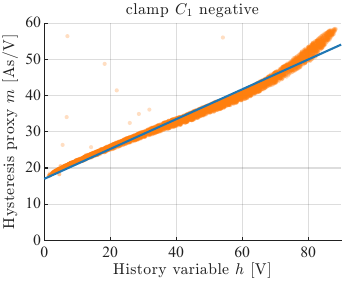}
    \end{minipage}
    \begin{minipage}{0.32\textwidth}
        \centering
        \includegraphics[width=\textwidth]{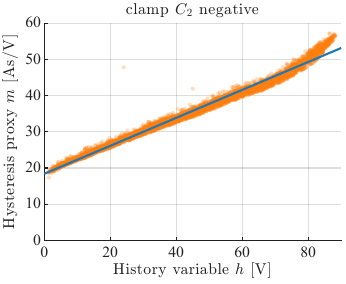}
    \end{minipage}
    \begin{minipage}{0.32\textwidth}
        \centering
        \includegraphics[width=\textwidth]{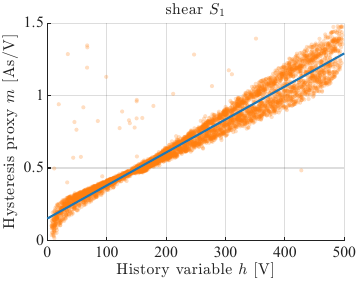}
    \end{minipage}
    \begin{minipage}{0.32\textwidth}
        \centering
        \includegraphics[width=\textwidth]{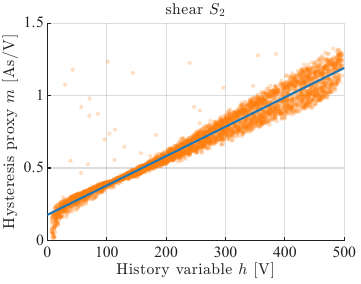}
    \end{minipage}
    \caption{Measured hysteresis proxy $m$~\eqref{eq:proxy} against the history variable $h$~\eqref{eq:history} for the clamp and shear elements, each pooling samples recorded over a range of drive frequencies~(\protect\legsym{mplorange}[none][circle]), together with the affine fit~\eqref{eq:hyst_model}~(\protect\legsym{mplblue}[solid]).}
    \label{fig:hyst_loops}
\end{figure*}

To identify the gain, each element is driven open-loop with steady-state voltage sweeps over drive frequencies from \SI{0.1}{\hertz} to \SI{50}{\hertz}, while its voltage and current are recorded. Figure~\ref{fig:hyst_loops} shows the resulting proxy $m$ against the history $h$. The measurements fall along a single affine trend in $h$, with $R^2$ from $0.944$ to $0.985$ for the clamps and from $0.946$ to $0.952$ for the shears, so the rate dependence of the gain is small.

We adopt this affine gain,
\begin{equation}\label{eq:hyst_model}
    \hat{M}(h) = \theta_1\,h + \theta_2,
\end{equation}
with intercept $\theta_2 > 0$, the gain just after a reversal, and slope $\theta_1 \geq 0$, the rate at which it grows with the excursion $h$. Because the sweeps span a range of voltage rates and still share this single trend, a rate-independent gain suffices here. This affine gain simplifies the rate-dependent model of~\cite{VanMeer2025a} by dropping the dependence on the voltage rate.

In discrete time, the element relation is
\begin{equation}\label{eq:hysteresis}
    y(k) - y(k{-}1) = M\!\bigl(h(k)\bigr)\,\bigl(u(k) - u(k{-}1)\bigr),
\end{equation}
with the history at sample $k$
\begin{equation}\label{eq:history}
    h(k) = \big|u(k) - u(\bar{k})\big|,
\end{equation}
and $\bar{k}$ the most recent reversal sample. The parameters of $\hat{M}$ are fitted to the proxy by least squares,
\begin{equation}\label{eq:hyst_cost}
    (\hat\theta_1, \hat\theta_2) = \arg\min_{(\theta_1,\,\theta_2)\,\in\,\mathbb{R}^2} \sum_{k=1}^{N_M} \bigl(m(k) - \theta_1\,h(k) - \theta_2\bigr)^2,
\end{equation}
over the $N_M$ recorded samples, leaving out those near a reversal where $\dot{u} \approx 0$ makes $m(k)$ unreliable. Following the clamp and shear split above, each clamp element is fitted separately for its two stepping directions and each shear element with a single parameter pair.

Inverting~\eqref{eq:hysteresis} and using the identified $\hat{M}$ in place of $M$ yields the feedforward control law
\begin{equation}\label{eq:controllaw}
    u(k) = u(k{-}1) + \frac{r(k) - r(k{-}1)}{\hat{M}\!\bigl(h(k{-}1)\bigr)},
\end{equation}
where $r(k)$ is the reference displacement from the (possibly ILC-modified) waveform. The history variable at the previous sample $h(k-1)$ is used in place of $h(k)$, since $h(k)$ depends on the voltage $u(k)$ before it is applied.

The proxy~\eqref{eq:proxy} identifies $\hat{M}$ only up to the unknown constant $\xi$, so the control law~\eqref{eq:controllaw} reaches the intended displacement up to that same scale. This scale is inconsequential in practice: the reference $r$ carries the same units, and its amplitude is set so that the applied voltage spans the full element range over a commutation step, which absorbs $\xi$ into the reference~\cite{VanMeer2025a}.

\subsection{ILC for $\alpha$-Domain Disturbance Compensation}\label{sec:ilc}
In~\cite{VanMeer2025a}, a single compensation function was learned for one PSA, using the mover position error from a capacitive sensor. This section extends the approach to three PSAs by learning compensation functions sequentially: each PSA is calibrated individually while the others remain stationary, so that the POI error can be unambiguously attributed to the active PSA. The POI error is obtained from the high-rate proxy $\hat{p}$ (Section~\ref{sec:enc_est}) and projected onto piezo coordinates through the inverse kinematics.

The ILC correction modifies the shear waveforms only; clamp waveforms control the contact timing and are not adjusted by ILC. Because only one shear element is in contact with the mover at any time, a single scalar compensation function simultaneously modifies both shear waveforms. Separate functions are learned for positive and negative stepping directions, giving six compensation functions in total across the three PSAs.

During calibration of PSA $i$, only that PSA moves while the other two remain stationary. The ILC uses the projected error $\varepsilon_i$ of~\eqref{eq:proj}, evaluated at the POI proxy $p_e = \hat{p}$. Because only PSA $i$ is moving, this error isolates the contribution of that PSA, while the errors in the other two coordinates remain near zero (aside from measurement noise and kinematic coupling) and are not used for learning. This sequential approach simplifies the calibration procedure and ensures that each compensation function addresses only the disturbances from its own PSA.

As in Section~\ref{sec:hysteresis}, the learning procedure is identical for every compensation function, so the equations below are written for a single generic case: PSA and direction indices are omitted and $\varepsilon$ denotes the projected error $\varepsilon_i$ for whichever PSA is under consideration. The ILC is performed in open-loop fashion. The piezo walks at a constant rate through the commutation cycle, and the measurements are used only to update the feedforward compensation for the next trial.

The modified waveform is
\begin{equation}\label{eq:ff}
    \tilde{\rho}(\alpha) = \rho(\alpha) + f(\alpha),
\end{equation}
where $\rho$ is the nominal waveform and $f$ is the learned correction. Because the correction is parameterized in $\alpha$ rather than in time, a correction learned at one stepping speed applies at any other speed for which the stage behaves quasi-statically. When these modified waveforms are used together with the hysteresis compensation of Section~\ref{sec:hysteresis}, the resulting element displacements approximately track the modified references and the mover follows the intended trajectory more accurately.

The update law for trial $j$ reads
\begin{equation}\label{eq:ilc}
    f_{j+1}(k) = Q(z)\big(f_{j}(k) + L(z)\,\varepsilon_{j}(k)\big),
\end{equation}
with robustness filter $Q$ and learning filter $L$, and where $G$ denotes the plant transfer from element waveform modification $f$ to the projected PSA error $\varepsilon$. Substituting~\eqref{eq:ilc} into the system yields the trial-to-trial error dynamics $\varepsilon_{j+1} = Q(1 - LG)\,\varepsilon_j$~\cite{Bristow2006}. With $L = G^{-1}$ and $Q = 1$, the factor $(1 - LG)$ vanishes, so $\varepsilon_{j+1} = 0$ regardless of the initial error and perfect compensation is achieved in a single trial. In practice, however, $G$ is not perfectly known and $L$ can only approximate $G^{-1}$, so that $Q$ must be designed to ensure convergence despite the model mismatch.

A sufficient condition for monotonic convergence of~\eqref{eq:ilc} is~\cite{Bristow2006}
\begin{equation}\label{eq:converge}
    \sup_{\omega \in [0,\pi]} \big|Q(e^{j\omega})\big(1 - L(e^{j\omega})\, G(e^{j\omega})\big)\big| < 1.
\end{equation}
The learning filter is set to $L = \hat{G}^{-1}$, where $\hat{G}$ is a parametric model of $G$ (identified below). Ideally, $|Q|$ should be close to 1 everywhere to retain the full learned correction. However, at frequencies where $\hat{G}$ deviates from $G$, the factor $(1 - LG)$ is no longer small and $|Q|=1$ can violate~\eqref{eq:converge}. In many applications, the model is most accurate at low frequencies, so $Q$ is designed as a low-pass Butterworth filter. The convergence condition is evaluated on a non-parametric frequency response $\bar{G}(e^{j\omega})$ obtained from measured data (see below) to more closely reflect the true system behavior.

The shear element plant model is identified following~\cite{VanMeer2025a} using a best-linear-approximation (BLA) procedure~\cite{Pintelon2012}: the commutation angle $\alpha$ is perturbed by a random-phase multisine signal, and the resulting displacement is measured via the corrected encoder estimate $\hat{p}$. The BLA extracts a non-parametric frequency response $\bar{G}(e^{j\omega})$ from this input--output data, to which a low-order parametric model $\hat{G}$ is fitted.

\begin{remark}
The ILC operates in piezo coordinates, since both the projected error $\varepsilon_i$ and the waveform correction $f$ live in the $q$-domain. The plant model $\hat{G}$ must therefore be identified consistently with the sensor-deviation-corrected output used for learning. The recorded output is the projected corrected estimate $[\Gamma(\hat{p},\varphi)]_i$ rather than the raw encoder reading $q_i$, so that $\hat{G}$ captures the transfer from waveform modification to projected corrected error. The two outputs differ by $\Delta_i(\alpha_i)$, a static function of the commutation angle $\alpha$, which is the excitation variable. A static (nonlinear) map contributes at most a real, frequency-independent constant to the best linear approximation, and its remaining effect is a stochastic nonlinear distortion that only raises the variance of $\bar{G}$~\cite{Pintelon2012}, so $\Delta_i$ introduces no frequency-dependent dynamics into $\hat{G}$. Since the same corrected output is used both for identifying $\hat{G}$ and for running the ILC, this constant is consistently absorbed into $L = \hat{G}^{-1}$, and the convergence condition~\eqref{eq:converge} is unaffected.
\end{remark}

\begin{figure}[t!]
    \centering
    \includegraphics[width=0.95\linewidth]{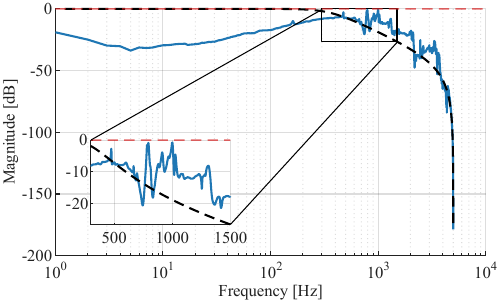}
    \caption{Verification of the ILC convergence condition~\eqref{eq:converge} for the representative PSA. Shown are $|Q(1 - L\bar{G})|$ (\protect\legsym{mplblue}[solid]), $|Q|$ (\protect\legsym{black}[dashed]), and the 0~dB threshold (\protect\legsym{mplred}[dashed]). The inset zooms the region around 1~kHz where the margin is smallest; the condition is satisfied across the full frequency range.}
    \label{fig:ilc_condition}
\end{figure}

The time-domain signal $f_{j+1}(k)$ from~\eqref{eq:ilc} must be converted to a function of $\alpha$. The compensation function is parameterized as a linear combination of basis functions,
\begin{equation}\label{eq:basis}
    f(\alpha) = \boldsymbol{\psi}_f^\top(\alpha)\, \boldsymbol{\gamma},
\end{equation}
where $\boldsymbol{\psi}_f(\alpha) \in \mathbb{R}^{n_f}$ collects $n_f$ basis functions and $\boldsymbol{\gamma} \in \mathbb{R}^{n_f}$ are the coefficients. In practice, piecewise-linear basis functions on $n_f$ equidistant grid points in $[0, 2\pi)$ are used. After each ILC update~\eqref{eq:ilc}, the time-domain signal is projected onto this basis by minimizing the weighted squared residual
\begin{equation}\label{eq:ilc_cost}
    \boldsymbol{\gamma}_{j+1} = \arg\min_{\boldsymbol{\gamma} \in \mathbb{R}^{n_f}} \big\| \hat{\mathbf{G}} (\boldsymbol{\Psi}_f \boldsymbol{\gamma} - \mathbf{f}_{j+1}) \big\|_2^2,
\end{equation}
where $\boldsymbol{\Psi}_f = [\boldsymbol{\psi}_f(\alpha(1)), \ldots, \boldsymbol{\psi}_f(\alpha(N))]^\top$ stacks the basis function evaluations, $\hat{\mathbf{G}}$ is the lifted impulse response matrix of $\hat{G}$, and $\mathbf{f}_{j+1}$ stacks the time-domain signal $f_{j+1}(k)$, see~\cite{VanMeer2025a}. The minimization has the same structure as~\eqref{eq:lsq_solution}, with the solution
\begin{equation}\label{eq:ilcproj}
    \boldsymbol{\gamma}_{j+1} = (\boldsymbol{\Psi}_f^\top \hat{\mathbf{G}}^\top \hat{\mathbf{G}} \boldsymbol{\Psi}_f)^{-1} \boldsymbol{\Psi}_f^\top \hat{\mathbf{G}}^\top \hat{\mathbf{G}}\, \mathbf{f}_{j+1}.
\end{equation}
The premultiplication by $\hat{\mathbf{G}}$ ensures that the projection minimizes the predicted output error rather than the compensation signal itself. The basis parameterization and projection modify the convergence analysis. The exact necessary and sufficient condition for monotonic convergence of the combined scheme~\eqref{eq:ilc}--\eqref{eq:ilcproj} involves both $\boldsymbol{\Psi}_f$ and $\hat{\mathbf{G}}$~\cite[Lemma~4.1]{VanMeer2025a},~\cite{Boeren2016a}. However,~\cite[Theorem~4.2]{VanMeer2025a} shows that the frequency-domain condition~\eqref{eq:converge} is sufficient regardless of the basis choice, so that condition is used here for filter design. Figure~\ref{fig:ilc_condition} verifies~\eqref{eq:converge} on the measured FRF $\bar{G}$ of each PSA, shown here for the representative one. The cutoff of $Q$ is designed to be as high as possible while maintaining $|Q(1 - L\bar{G})| < 1$.

\section{Experimental Results}\label{sec:experiments}
This section evaluates the framework experimentally, first on the lab setup and then on the operational electron microscope. All displacement quantities in the results below are reported in arbitrary units (a.u.), obtained through a single fixed scaling applied consistently across every figure and value.

Throughout, the tracking error is evaluated in piezo coordinates. Only one PSA is stepped at a time, so its error is a single scalar, obtained by projecting a position signal onto that PSA's coordinate through the inverse kinematics and subtracting the piezo reference, as in~\eqref{eq:proj}. We report this scalar for three position signals, namely the raw encoder estimate $\hat{p}_0$, the encoder estimate after sensor-deviation correction $\hat{p}$, and the true POI measurement given by the capacitive reading $p_c$ on the lab setup and the vision-based estimate $p_s$ on the microscope. On the lab setup, the framework was calibrated on all three PSAs and the improvements are comparable across them. The results below therefore focus on one representative PSA on both the lab setup and the microscope. On the EM, its axis is aligned with the imaging plane by choice of tilt angle (Section~\ref{sec:vision}), so the in-plane image shift observes the full motion of that axis.

% --- ILC CONVERGENCE ---
\begin{figure}[t!]
    \centering
    \includegraphics[width=\linewidth]{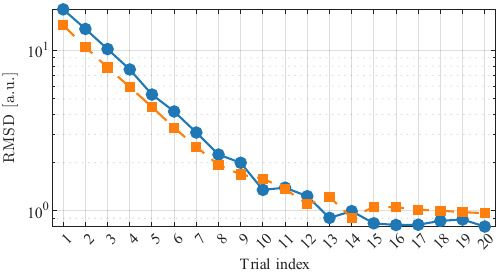}
    \caption{RMSD of the projected tracking error $\varepsilon_i$~\eqref{eq:proj} versus ILC trial index. Forward stepping (\protect\legsym{mplblue}[solid][circle]) and backward stepping (\protect\legsym{mplorange}[dashed][square]) are overlaid. Hysteresis compensation is active in all trials.}
    \label{fig:ilc_convergence}
\end{figure}

Four feedforward strategies are compared throughout the experiments. The difference between them lies in which compensators are active:
\begin{itemize}[nosep]
    \item[\textcolor{mplblue}{\textbf{S1}}] Traditional feedforward with a constant (non-compensated) hysteresis model and the unmodified nominal waveforms $\rho$, i.e., $\hat{M}(h) = 1$, $\Delta(\alpha) = 0$, and $f(\alpha) = 0$.
    \item[\textcolor{mplorange}{\textbf{S2}}] Rate-independent hysteresis compensation only (Section~\ref{sec:hysteresis}), with $\Delta(\alpha) = 0$ and $f(\alpha) = 0$.
    \item[\textcolor{mplgreen}{\textbf{S3}}] Hysteresis compensation and ILC, without sensor deviation correction: $\Delta(\alpha) = 0$.
    \item[\textcolor{mplred}{\textbf{S4}}] Full framework: hysteresis compensation, sensor deviation correction, and ILC.
\end{itemize}

\subsection{ILC Convergence}
The ILC compensation is learned at a drive frequency of \SI{2}{\hertz}, on the lab setup for each PSA in turn and on the EM for the representative PSA. Because the correction is parameterized in $\alpha$, it is subsequently applied across the full tested frequency range. Figure~\ref{fig:ilc_convergence} shows the RMSD of the projected tracking error $\varepsilon_i$~\eqref{eq:proj} over successive ILC trials for the representative PSA on the lab setup, in both stepping directions. Hysteresis compensation is active in all trials.

\subsection{Positioning Performance on the Lab Setup}
\begin{figure*}[t!]
    \centering
    \begin{minipage}{0.32\textwidth}
        \centering
        \includegraphics[width=\textwidth]{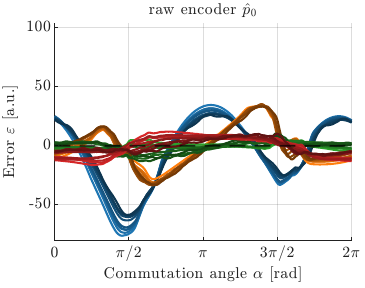}
    \end{minipage}
    \begin{minipage}{0.32\textwidth}
        \centering
        \includegraphics[width=\textwidth]{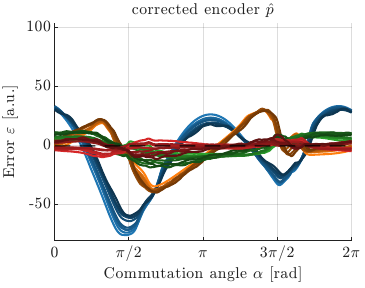}
    \end{minipage}
    \begin{minipage}{0.32\textwidth}
        \centering
        \includegraphics[width=\textwidth]{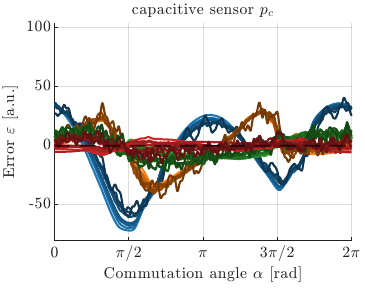}
    \end{minipage}
    \caption{Mean projected tracking error $\varepsilon_i$~\eqref{eq:proj} versus commutation angle $\alpha_i$ on the lab setup (positive stepping direction). Left to right: evaluated from $\hat{p}_0$ (raw encoder), $\hat{p}$ (sensor-deviation-corrected encoder), and $p_c$ (capacitive POI). Each panel shows strategies S1~(\protect\legsym{mplblue}[solid]), S2~(\protect\legsym{mplorange}[solid]), S3~(\protect\legsym{mplgreen}[solid]), and S4~(\protect\legsym{mplred}[solid]), with lines darkening from 0.1~Hz (lightest) to 10~Hz (darkest).}
    \label{fig:error_vs_time}
\end{figure*}

\begin{figure}[t!]
    \centering
    \begin{minipage}{0.155\textwidth}
        \centering
        \includegraphics[width=\textwidth]{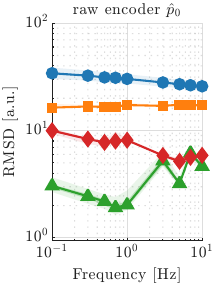}
    \end{minipage}
    \begin{minipage}{0.155\textwidth}
        \centering
        \includegraphics[width=\textwidth]{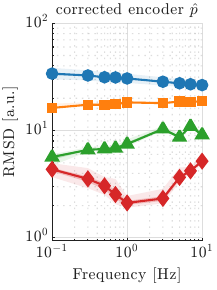}
    \end{minipage}
    \begin{minipage}{0.155\textwidth}
        \centering
        \includegraphics[width=\textwidth]{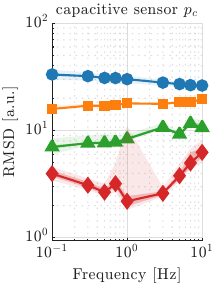}
    \end{minipage}
    \begin{minipage}{0.155\textwidth}
        \centering
        \includegraphics[width=\textwidth]{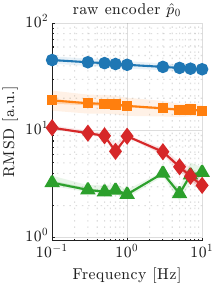}
    \end{minipage}
    \begin{minipage}{0.155\textwidth}
        \centering
        \includegraphics[width=\textwidth]{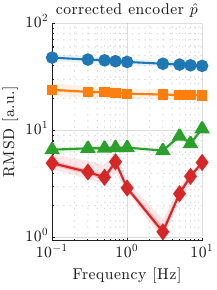}
    \end{minipage}
    \begin{minipage}{0.155\textwidth}
        \centering
        \includegraphics[width=\textwidth]{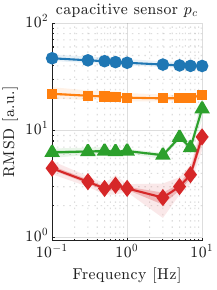}
    \end{minipage}
    \caption{RMSD of the projected tracking error $\varepsilon_i$~\eqref{eq:proj} versus drive frequency: S1~(\protect\legsym{mplblue}[solid][circle]), S2~(\protect\legsym{mplorange}[solid][square]), S3~(\protect\legsym{mplgreen}[solid][triangle]), S4~(\protect\legsym{mplred}[solid][diamond]); shaded bands of matching color denote the IQR and 5th--95th percentile range over repeated steps. Top row: positive stepping; bottom row: negative stepping. Within each row, the panels correspond from left to right to $\hat{p}_0$, $\hat{p}$, and $p_c$.}
    \label{fig:rmsd_vs_freq}
\end{figure}

Figure~\ref{fig:error_vs_time} compares the POI tracking error versus commutation angle for strategies S1, S2, S3, and S4 on the lab setup at drive frequencies from \SI{0.1}{\hertz} to \SI{10}{\hertz}. Figure~\ref{fig:rmsd_vs_freq} summarizes the error per drive frequency.

% --- BEFORE/AFTER ERROR VS TIME (lab) ---

% --- BEFORE/AFTER ERROR VS DRIVE FREQUENCY (lab) ---

From the figures, each feedforward component contributes to improved POI tracking. At a drive frequency of \SI{1}{\hertz}, the POI RMSD measured by the capacitive sensor drops from \SI{35.5}{\arbunit} for the traditional strategy S1 to \SI{2.7}{\arbunit} for the full framework S4 in the positive stepping direction, and from \SI{48.5}{\arbunit} to \SI{3.5}{\arbunit} in the negative direction. These are 13- and 14-fold reductions, respectively, and the same trend holds across the tested drive frequencies (Figure~\ref{fig:rmsd_vs_freq}). At the highest tested drive frequencies, the performance of S3 and S4 drops off slightly, which is mostly attributed to flexible-mode behavior of the stage that the quasi-static feedforward does not target.

Figure~\ref{fig:error_vs_time} also shows that the raw encoder error for strategy S4 is worse than for S3. This is by design: the sensor deviation correction in combination with ILC results in a shaped non-zero error at the encoder level, but it is precisely this shaped error that yields reduced error at the POI, which is the actual control objective.

\subsection{Positioning Performance on the EM}
The lab-setup results above confirm the feedforward components (hysteresis compensation, sensor deviation, and ILC) using capacitive ground truth at the POI. On the EM, the feedforward models are re-identified following the same procedures as on the lab setup, with hysteresis from element voltage and current, sensor deviation from $p_s$, and ILC compensation from the corrected encoder estimate $\hat{p}$.

\begin{figure*}[t!]
    \centering
    \begin{minipage}{0.32\textwidth}
        \centering
        \includegraphics[width=\textwidth]{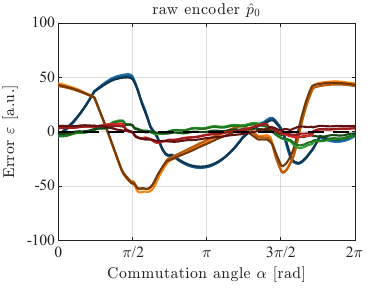}
    \end{minipage}
    \begin{minipage}{0.32\textwidth}
        \centering
        \includegraphics[width=\textwidth]{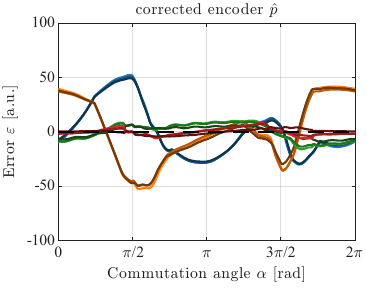}
    \end{minipage}
    \begin{minipage}{0.32\textwidth}
        \centering
        \includegraphics[width=\textwidth]{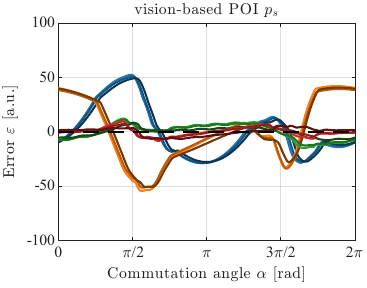}
    \end{minipage}
    \caption{Mean projected tracking error $\varepsilon_i$~\eqref{eq:proj} versus commutation angle $\alpha_i$ on the EM stage (positive stepping direction). Left to right: evaluated from $\hat{p}_0$ (raw encoder), $\hat{p}$ (sensor-deviation-corrected encoder), and $p_s$ (vision-based POI). Each panel shows strategies S1~(\protect\legsym{mplblue}[solid]), S2~(\protect\legsym{mplorange}[solid]), S3~(\protect\legsym{mplgreen}[solid]), and S4~(\protect\legsym{mplred}[solid]), with lines darkening from 0.5~Hz (lightest) to 2~Hz (darkest).}
    \label{fig:em_error_alpha}
\end{figure*}

\begin{figure}[t!]
    \centering
    \begin{minipage}{0.155\textwidth}
        \centering
        \includegraphics[width=\textwidth]{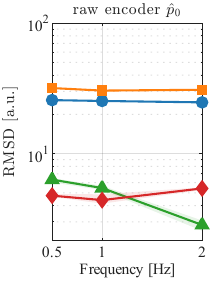}
    \end{minipage}
    \begin{minipage}{0.155\textwidth}
        \centering
        \includegraphics[width=\textwidth]{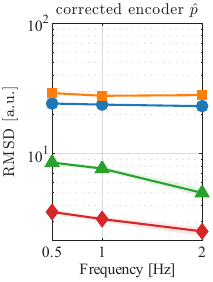}
    \end{minipage}
    \begin{minipage}{0.155\textwidth}
        \centering
        \includegraphics[width=\textwidth]{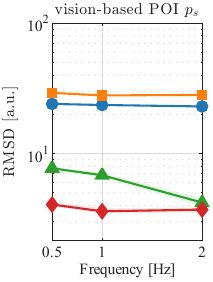}
    \end{minipage}
    \begin{minipage}{0.155\textwidth}
        \centering
        \includegraphics[width=\textwidth]{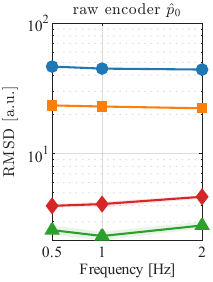}
    \end{minipage}
    \begin{minipage}{0.155\textwidth}
        \centering
        \includegraphics[width=\textwidth]{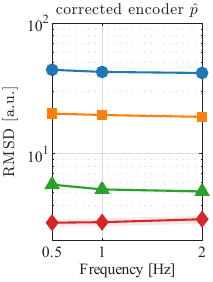}
    \end{minipage}
    \begin{minipage}{0.155\textwidth}
        \centering
        \includegraphics[width=\textwidth]{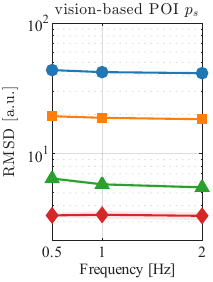}
    \end{minipage}
    \caption{RMSD of the projected tracking error $\varepsilon_i$~\eqref{eq:proj} versus drive frequency on the EM stage: S1~(\protect\legsym{mplblue}[solid][circle]), S2~(\protect\legsym{mplorange}[solid][square]), S3~(\protect\legsym{mplgreen}[solid][triangle]), S4~(\protect\legsym{mplred}[solid][diamond]); shaded bands of matching color denote the IQR and 5th--95th percentile range over repeated steps. Top row: positive stepping; bottom row: negative stepping. Within each row, the panels correspond from left to right to $\hat{p}_0$, $\hat{p}$, and $p_s$.}
    \label{fig:em_rmsd_vs_freq}
\end{figure}

Figure~\ref{fig:em_error_alpha} shows the projected tracking error $\varepsilon_i$ as a function of commutation angle $\alpha_i$ for a representative PSA on the EM stage at drive frequencies from \SI{0.5}{\hertz} to \SI{2}{\hertz}, analogous to Figure~\ref{fig:error_vs_time} for the lab setup. The residual structure on the EM mirrors that on the lab setup, confirming that the same physical effects are present and that the modeling approach transfers to the operational microscope. Figure~\ref{fig:em_rmsd_vs_freq} summarizes the EM tracking error versus drive frequency, mirroring Figure~\ref{fig:rmsd_vs_freq} for the lab setup. At \SI{1}{\hertz}, the full framework S4 reduces the POI RMSD from \SI{32.7}{\arbunit} to \SI{4.7}{\arbunit} in the positive stepping direction, and from \SI{55.2}{\arbunit} to \SI{4.7}{\arbunit} in the negative direction, a 7-fold and 12-fold reduction. The improvement is consistent across the tested drive frequencies. We observe a slight performance drop compared to the lab setup, only in the ILC contribution. This is attributed to position-dependent stage dynamics: the ILC compensation is learned at the stage home position, whereas the vision-based validation requires imaging the specimen at a different stage position.

% --- EM COMMUTATION-ANGLE ERROR ---

% --- EM RMSD vs drive frequency ---

\section{Conclusion}\label{sec:conclusion}
This paper presented an integrated framework for precision specimen positioning in an operational electron microscope. Positioning accuracy in such instruments is limited by hysteresis, mechanical misalignments, non-collocated sensing, and non-repeatable disturbances such as thermal drift. This work builds upon prior work that has addressed several of these challenges on a simplified lab setup~\cite{VanMeer2025a}. In this work the framework is extended to operation in a real transmission electron microscope.

The gap between encoder and POI coordinates is bridged by a commutation-angle-dependent sensor deviation that is physically motivated by bending from clamp-shear contact forces. The deviation is learned from whichever ground-truth sensor is available: capacitive sensors $p_c$ on the lab setup, accumulated image-tracker shifts $p_s$ on the EM. The corrected encoder estimate $\hat{p}$ provides a high-rate POI proxy that drives feedforward learning in both environments. The $\alpha$-domain ILC is extended from one PSA to three by projecting POI errors onto per-PSA signals through the inverse kinematics $\Gamma$. The hysteresis model is simplified from the rate-dependent radial-basis-function formulation used in~\cite{VanMeer2025a}. The new model is an affine function of the history variable, which reduces the parameter count from the number of radial basis functions to two per element.

On the lab setup, the framework reduces the POI RMSD 13-fold (positive stepping) and 14-fold (negative stepping) at \SI{1}{\hertz}, and on the electron microscope 7-fold and 12-fold, respectively. The improvement is consistent across the tested drive frequency range, though at higher drive frequencies performance slightly degrades, likely due to flexible-mode behavior. Importantly, this accuracy is reached with no sensor at the POI, since the corrections are learned from the images that the microscope already acquires. The framework therefore enables nanometer-scale specimen positioning without added sensing hardware.

Several directions remain for future work. Firstly, the performance gain during tomographic tilt series should be validated in practice. Next, combining the present feedforward framework with the image-based feedback of~\cite{VanHorssen2020} would let the two together reject both the repeatable and the non-repeatable disturbances. This becomes more powerful once the image and encoder measurements are fused: the low-level feedback could then be driven by a fused, high-rate POI estimate rather than the encoder signal alone, which has the potential to substantially improve closed-loop performance. Estimating the out-of-plane coordinate $p_z$ from image defocus would extend the vision sensor to three dimensions. Finally, flexible-mode vibrations are repeatable but not $\alpha$-periodic, so the current feedforward does not address them; time-domain approaches such as input shaping could target these residuals.

\bibliographystyle{IEEEtran}
\bibliography{references}
\end{document}